\documentclass{svproc}
\usepackage{graphicx}
\usepackage{multirow}
\usepackage{amsmath,amssymb,amsfonts}
\usepackage{mathrsfs}
\usepackage[title]{appendix}
\usepackage{xcolor}
\usepackage{textcomp}
\usepackage{manyfoot}
\usepackage{booktabs}
\usepackage{algorithm}
\usepackage{algorithmicx}
\usepackage{algpseudocode}
\usepackage{listings}
\usepackage{url}

\usepackage{booktabs,tabularx,siunitx,threeparttable}

\begin{document}
\mainmatter

\title{Who Coordinates U.S. Cyber Defense? A Co-Authorship Network Analysis of Joint Cybersecurity Advisories (2024--2025)}
\titlerunning{Co-Author Network of Cybersecurity Advisories}

\author{M. Abdullah Canbaz\inst{1} \and Hakan Otal\inst{1} \and Tugce Unlu\inst{1} \and Nour Alhussein\inst{2} \and Brian Nussbaum\inst{2}}

\institute{Department of Information Sciences and Technology, \\
\email{mcanbaz, hotal, tunlu [at] albany \{dot\} edu}
\and
Department of Cybersecurity, \\
\email{nalhussein, bnussbaum [at] albany \{dot\} edu}
\\College of Emergency Preparedness, Homeland Security, and Cybersecurity, \\
University at Albany, SUNY, Albany, NY, United States \\
}

\maketitle

\begin{abstract}
Cyber threats increasingly demand joint responses, yet the organizational dynamics behind multi-agency cybersecurity collaboration remain poorly understood. Understanding who leads, who bridges, and how agencies coordinate is critical for strengthening both U.S. homeland security and allied defense efforts. In this study, we construct a co-authorship network from nine Joint Cybersecurity Advisories (CSAs) issued between November 2024 and August 2025. We map 41 agencies and 442 co-authoring ties to analyze the structure of collaboration. We find a tightly knit U.S. triad—CISA, FBI, and NSA—densely connected with Five Eyes and select European allies. Degree centrality identifies CISA and FBI as coordination hubs, while betweenness highlights NSA, the UK’s NCSC, and Australia’s ASD-ACSC as key bridges linking otherwise fragmented clusters. By releasing the first replicable dataset and network analysis of CSAs, we provide new empirical evidence on how collaborative cybersecurity signals are organized and where strategic influence is concentrated.
\end{abstract}

\section{Introduction}
Homeland security increasingly hinges on the ability to anticipate, detect, and mitigate cyber threats through multinational and interagency collaboration. While public advisories issued by U.S. and allied agencies provide tactical details, what remains opaque is the \emph{collaboration topology} behind them: 
who coordinates, which agencies consistently appear together, and how the coalition structures itself across threat domains. 

The study of national cyber defense as a networked system is essential for two reasons. 
First, cyber threats exhibit \emph{complex network dynamics} themselves---botnets, ransomware affiliates, and fast-flux DNS infrastructures are inherently network-structured \cite{chen2020botnets,dittrich2012internet}. 
Understanding defense as a counter-network provides an analytic symmetry: networks fight networks. 
Second, organizational resilience depends not only on technical capacity but also on the \emph{connectivity patterns} among agencies. 
In network science, metrics such as centrality, clustering, and k-core structure have been shown to reveal vulnerabilities and robustness in social, biological, and infrastructure systems \cite{newman2010networks,barabasi2016network}. 
Applying these methods to cyber defense collaboration can expose hidden strengths and potential single points of failure.

Similar work has examined information-sharing partnerships, such as ISACs (Information Sharing and Analysis Centers) \cite{goodman2010cyber}, public–private cyber alliances \cite{yadav2021information}, and international intelligence cooperation \cite{bigo2014intelligence}. 
These studies emphasize governance and policy dimensions, but few have mapped the \emph{empirical co-authorship structure} of joint cybersecurity advisories (CSAs). 
Recent computational social science research demonstrates the value of co-authorship network analysis for identifying intellectual leadership and collaboration clusters in science and technology domains \cite{hou2019coauthor,liu2020scientific}. 
We extend this approach to the domain of homeland security cyber defense, treating joint advisories as the observable “co-authored documents” of operational cooperation.

While public cyber advisories issued by the United States government have been studied \cite{lanz2022cybersecurity}, such studies have mostly addressed the content of such advisories rather than the co-authorship networks behind them; this analysis will expand understandings of these documents and the role they play in coordinating efforts across national boundaries.

\textbf{In this work}, we construct a new co-authorship network from nine multi-agency CSAs released between November 2024 and August 2025. 
We then apply network science methods---including centrality, clustering, assortativity, modularity, and k-core decomposition---to quantify the coalition’s structure. 
Our contribution is twofold: (1) we provide the first network-science based mapping of CSA authorship to reveal the structural nucleus of U.S. and allied cyber defense, and (2) we release the underlying edge and node tables for replication and further analysis. For clarity, we use the following abbreviations throughout: CSA (Joint Cybersecurity Advisory), CISA (Cybersecurity and Infrastructure Security Agency), ISAC (Information Sharing and Analysis Center), and “agency” to denote any national or sectoral institution formally co-authoring an advisory.

Our analysis is guided by the following research questions:
\begin{enumerate}
    \item Who are the structural hubs and brokers coordinating multinational cyber defense advisories?
    \item What are the meso-scale structures (communities, k-cores) that sustain resilience in the coalition?
    \item To what extent does the CSA network exhibit small-world and hub-centric properties typical of resilient networks?
\end{enumerate}

By answering these questions, we aim to shed light on how the institutional fabric of homeland security cyber defense is woven, and what network features enable it to scale against adversaries that themselves operate as distributed, adaptive networks.

\vspace{-2mm}
\section{Methodology}
\vspace{-2mm}
Our empirical basis is a corpus of nine Joint Cybersecurity Advisories (CSAs) issued between November 2024 and August 2025. These advisories were retrieved directly from the CISA public advisory archive and verified against mirrored postings on FBI and NSA portals. Each CSA lists its co-authoring institutions, typically including U.S. federal agencies, Five Eyes intelligence partners, European security services, and sector-specific regulators. For example, CSA-008 was co-signed by CISA, the FBI, the Australian Cyber Security Centre, and the UK’s National Cyber Security Centre, while CSA-004 featured contributions from the NSA, the Canadian Centre for Cyber Security, and Germany’s BSI. By extracting the set of institutional co-signatories from each advisory, we obtain a relational record of collaboration events.
\vspace{-2mm}
\subsection*{Data Collection}
\vspace{-2mm}
We collected the advisories through structured scraping and manual validation to ensure accuracy. For each advisory, we recorded its identifier (e.g., AA25-239A), publication date, threat type (APT, ransomware, infrastructure exploitation), and the complete set of authoring agencies.

Between November 2024 and August 2025, CISA issued a total of 14 Joint Cybersecurity Advisories (CSAs). We selected nine of these for detailed analysis. This period was chosen as it represents the first complete year in which multi-agency CSAs were consistently published in a harmonized format; while brief, it captures diverse adversary types. We note that extending the window may alter structural metrics, which we flag as a direction for future replication. The selection criterion was diversity of threat type and co-author composition: our final set spans state-sponsored APT campaigns, financially motivated ransomware operations, and large-scale infrastructure exploitation incidents. This stratified approach ensured that the corpus did not over-represent a single adversary tactic or a narrow subset of institutional collaborations.

Non-substantive acknowledgments, such as private-sector contributors mentioned in footnotes, were excluded so that the network would reflect formal inter-agency authorship. Institutional names were normalized to canonical acronyms, and agencies with overlapping acronyms were disambiguated by adding country qualifiers (e.g., distinguishing between NCSC-UK and NCSC-NZ). This procedure yielded a dataset of forty-one distinct agencies across nine advisories.
Although threat type (APT, ransomware, infrastructure exploitation) was recorded for each advisory, in this paper it is used descriptively rather than analytically; future work will stratify results by threat type to examine whether different adversary classes elicit different coalition structures.
\vspace{-2mm}
\subsection*{Feature Engineering and Graph Construction}
\vspace{-2mm}
To represent the collaboration structure, we adopted the standard co-authorship convention used in scientometrics. 
Within each CSA, all authoring agencies were treated as fully interconnected, producing a clique among participants. This assumption follows a standard scientometric convention, where co-authorship implies joint engagement, and it allows us to capture the “all-to-all” structure characteristic of joint publications such as multi-agency advisories.
If $k$ agencies co-authored an advisory, $\binom{k}{2}$ undirected edges were generated. 
When two agencies appeared together in multiple advisories, the corresponding edge weight was incremented to reflect the number of repeated collaborations. 
The final product is a single undirected, weighted graph $G=(V,E,W)$ containing forty-one nodes and 442 edges, where edge weights $w_{ij}$ represent the frequency of joint authorship between agencies $i$ and $j$.
\vspace{-2mm}
\subsection*{Analytical Measures}
\vspace{-2mm}
We applied a suite of node-level and global network measures to characterize the coalition. 
Degree centrality was calculated to capture the number of unique partners for each agency, while strength (weighted degree) measured the cumulative intensity of repeated collaborations. 
Betweenness centrality, computed with Brandes’ algorithm, highlighted agencies that act as crucial information brokers between otherwise disconnected parts of the coalition.
Clustering coefficients were used to examine triadic closure at the node level and overall network cohesiveness. 
Assortativity measured the correlation of degrees across connected agencies, distinguishing hub-to-hub from hub-to-periphery linking patterns. 
Average path length was estimated within the largest connected component to assess the efficiency of information flow. 
Community detection was performed through greedy modularity maximization, identifying clusters of closely tied agencies, while k-core decomposition revealed the structural nucleus, identifying groups of agencies so interconnected that they would remain linked even if many other agencies were removed.

This pipeline transforms raw textual advisories into a reproducible network dataset, enabling quantitative assessment of coalition structure. 
Our analysis is based on the one-mode projection of agency–advisory participation; while standard in co-authorship studies, a bipartite (two-mode) analysis could further reveal whether certain advisories disproportionately drive the observed structure.

\vspace{-2mm}
\section{Analysis and Results}
\vspace{-2mm}

The CSA co-authorship network consists of 41 agencies connected by 442 weighted ties across nine advisories. 
Each advisory contributes a dense clique among its co-signatories, resulting in a single connected component that links U.S. federal agencies, Five Eyes intelligence partners, and several European services. 
Table~\ref{tab:global} summarizes the global properties of the network.

The global statistics reveal a coalition that is both tightly connected and hierarchically structured. 
The average clustering coefficient of 0.902 indicates that agencies form highly cohesive triads, reflecting repeated joint appearances across advisories. 
The short average path length of 1.46 confirms a small-world structure: almost every agency can be reached through one or two intermediaries. 
At the same time, the negative degree assortativity ($r \approx -0.246$) shows that hubs disproportionately connect to smaller actors, producing a hub-and-spoke topology anchored by a few central agencies. 
Community detection identifies three meso-scale groups, but modularity is modest ($Q \approx 0.19$), implying that collaboration is widely shared across national boundaries rather than siloed within country clusters.

\begin{table}[!t]
\centering
\caption{Global metrics on the CSA co-authorship network (giant component).}
\label{tab:global}
\begin{tabular}{lc}
\toprule
Metric & Value \\
\midrule
Nodes & 41 \\
Edges & 442 \\
Connected components & 1 \\
Average clustering coefficient & 0.902 \\
Degree assortativity & -0.246 \\
Average shortest path length & 1.461 \\
Communities (greedy modularity) & 3 \\
Modularity & 0.190 \\
\bottomrule
\end{tabular}
\end{table}
\vspace{-2mm}
\subsection*{Central Agencies and Collaboration Intensity}
\vspace{-2mm}
Table~\ref{tab:metrics} ranks the top agencies by degree, weighted degree, and betweenness centrality. 
CISA and FBI dominate all three measures, each co-authoring with nearly every other organization and doing so repeatedly, as shown by their weighted degree of 73. 
NSA, ASD-ACSC, and NCSC-UK follow closely, combining high connectivity with substantial brokerage roles. 
European partners such as Germany’s BfV, Canada’s CSIS and CCCS, and the Netherlands’ MIVD form the second tier of consistently present collaborators. 
These patterns underscore that the U.S. triad of CISA–FBI–NSA is the backbone of the coalition, while Five Eyes and select European agencies broaden the perimeter of joint defense.

\begin{table}[tb]
\centering
\footnotesize
\caption{Top 10 agencies by network centrality (sorted by Degree; ties broken by Weighted Degree).}
\label{tab:metrics}
\begin{tabularx}{0.95\linewidth}{
  >{\raggedright\arraybackslash}X
  S[table-format=2]
  S[table-format=2]
  S[table-format=1.3]
}
\toprule
\multicolumn{1}{l}{Agency} & \multicolumn{1}{c}{Degree} & \multicolumn{1}{c}{\shortstack{Weighted\\Degree}} & \multicolumn{1}{c}{Betweenness} \\
\midrule
CISA & 40 & 73 & 0.201 \\
FBI & 40 & 73 & 0.201 \\
NSA & 38 & 66 & 0.106 \\
ASD--ACSC & 38 & 64 & 0.106 \\
NCSC--UK & 38 & 64 & 0.106 \\
BfV (Germany) & 32 & 39 & 0.046 \\
CSIS (Canada) & 32 & 39 & 0.046 \\
MIVD (Netherlands) & 32 & 39 & 0.046 \\
CCCS (Canada) & 29 & 47 & 0.040 \\
NCSC--NZ & 29 & 41 & 0.040 \\
\bottomrule
\end{tabularx}
\end{table}

\begin{figure}[!t]
\centering
\includegraphics[width=0.9\textwidth]{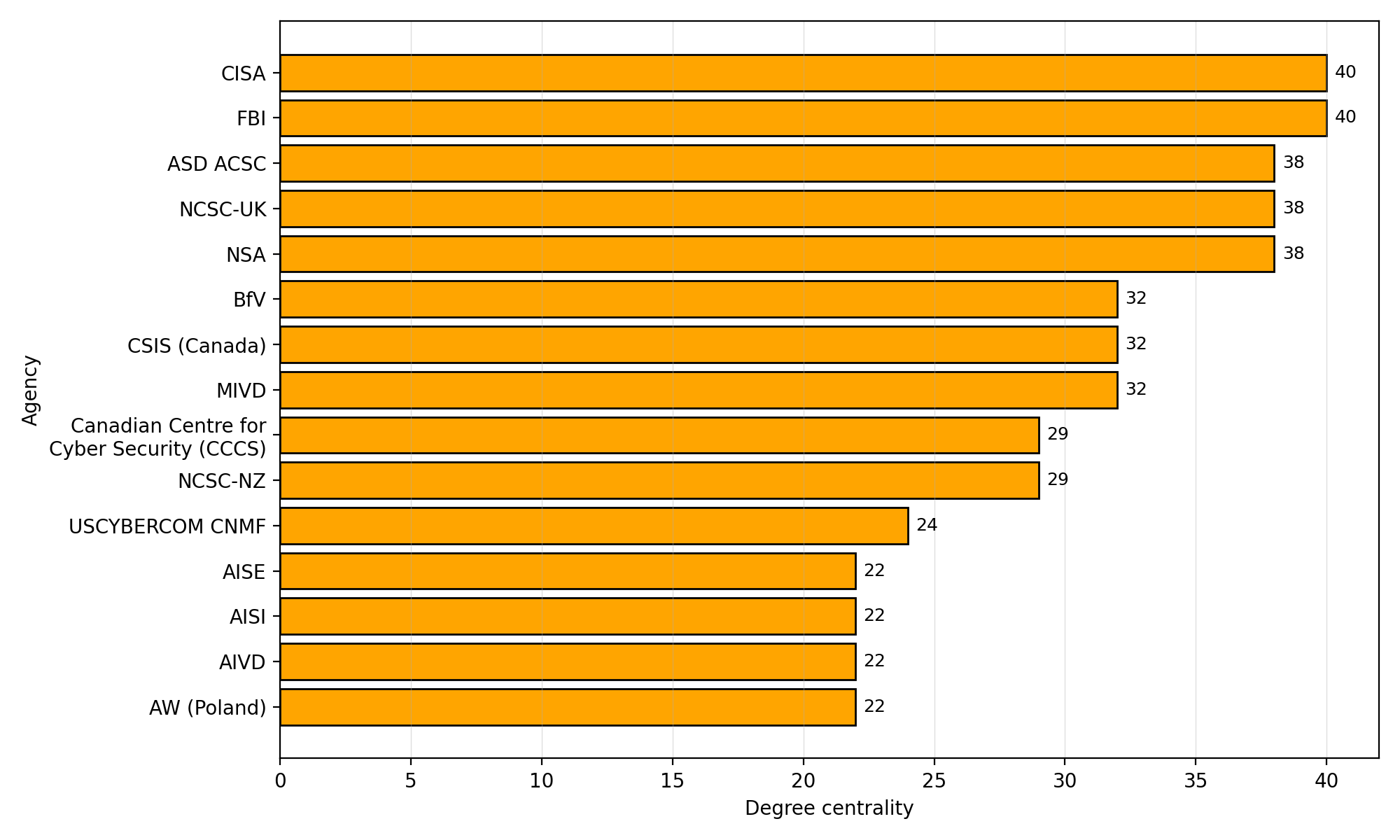}
\vspace{-6mm}
\caption{Top-15 agencies ranked by degree centrality.}
\vspace{-6mm}
\label{fig:top15}
\end{figure}

Figure~\ref{fig:top15} confirms this hierarchy by showing the top 15 agencies by co-author degree. 
The tallest bars belong to CISA and FBI, reflecting their nearly universal presence. 
The distribution then flattens toward the European second tier, showing that while participation is broad, only a handful of agencies consistently anchor every advisory.

\vspace{-2mm}
\subsection*{Network Structure and Collaboration Patterns}
\begin{figure}[ht]
\centering
\includegraphics[width=\textwidth]{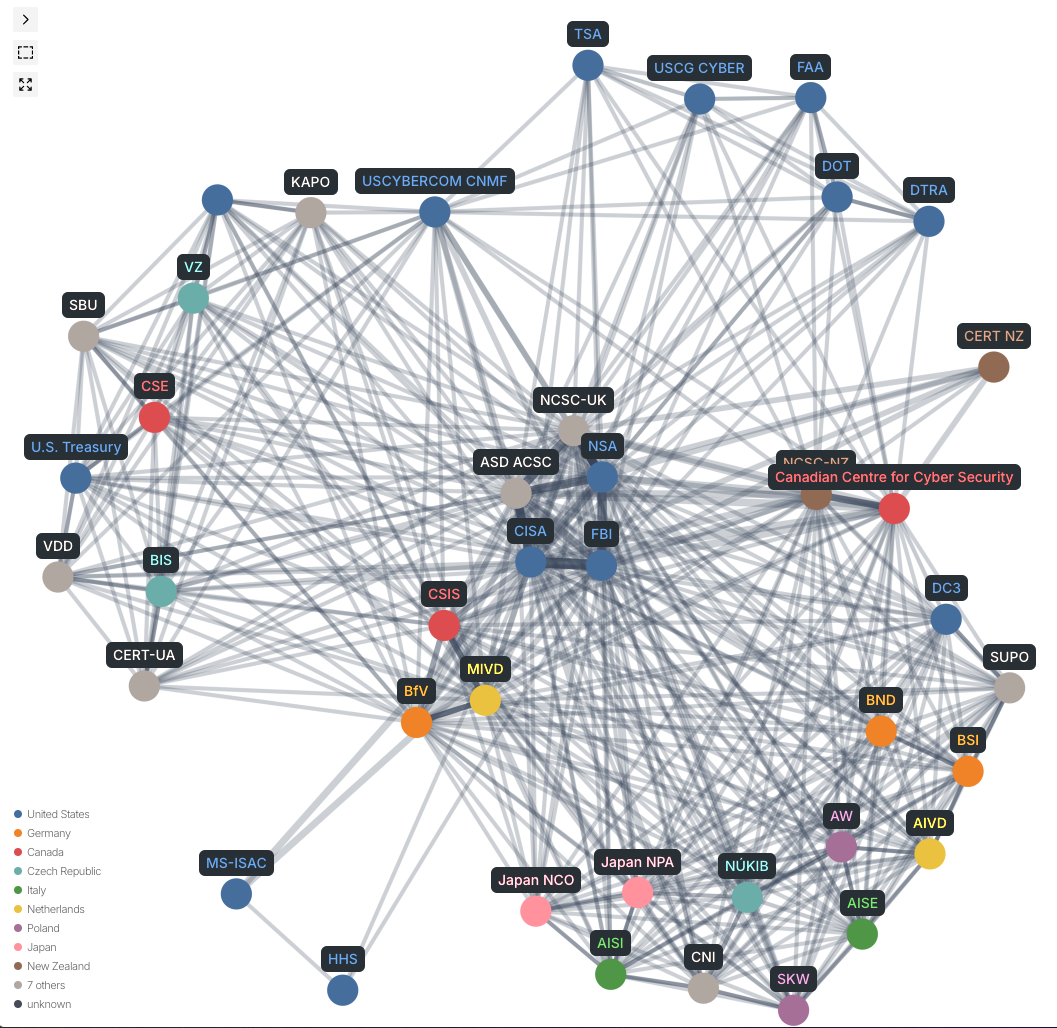}
\vspace{-6mm}
\caption{CSA co-authorship network. Edge width encodes frequency of joint authorship.}
\vspace{-6mm}
\label{fig:network}
\end{figure}

\vspace{-2mm}
The overall network can be visualized in Figure~\ref{fig:network}. 
The circular layout highlights the density of ties, but the variation in edge thickness reveals repeated collaborations among a subset of actors. 
The strongest dyads are CISA–FBI (9 co-authored advisories), FBI–NSA (6), and CISA–NSA (6). 
Other national agencies tend to connect to the network through one or more of these hubs, reinforcing the hub-and-spoke architecture observed in the metrics.

\vspace{-2mm}
\subsection*{Core–Periphery Patterns}
\begin{figure}[ht]
\centering
\includegraphics[width=\textwidth]{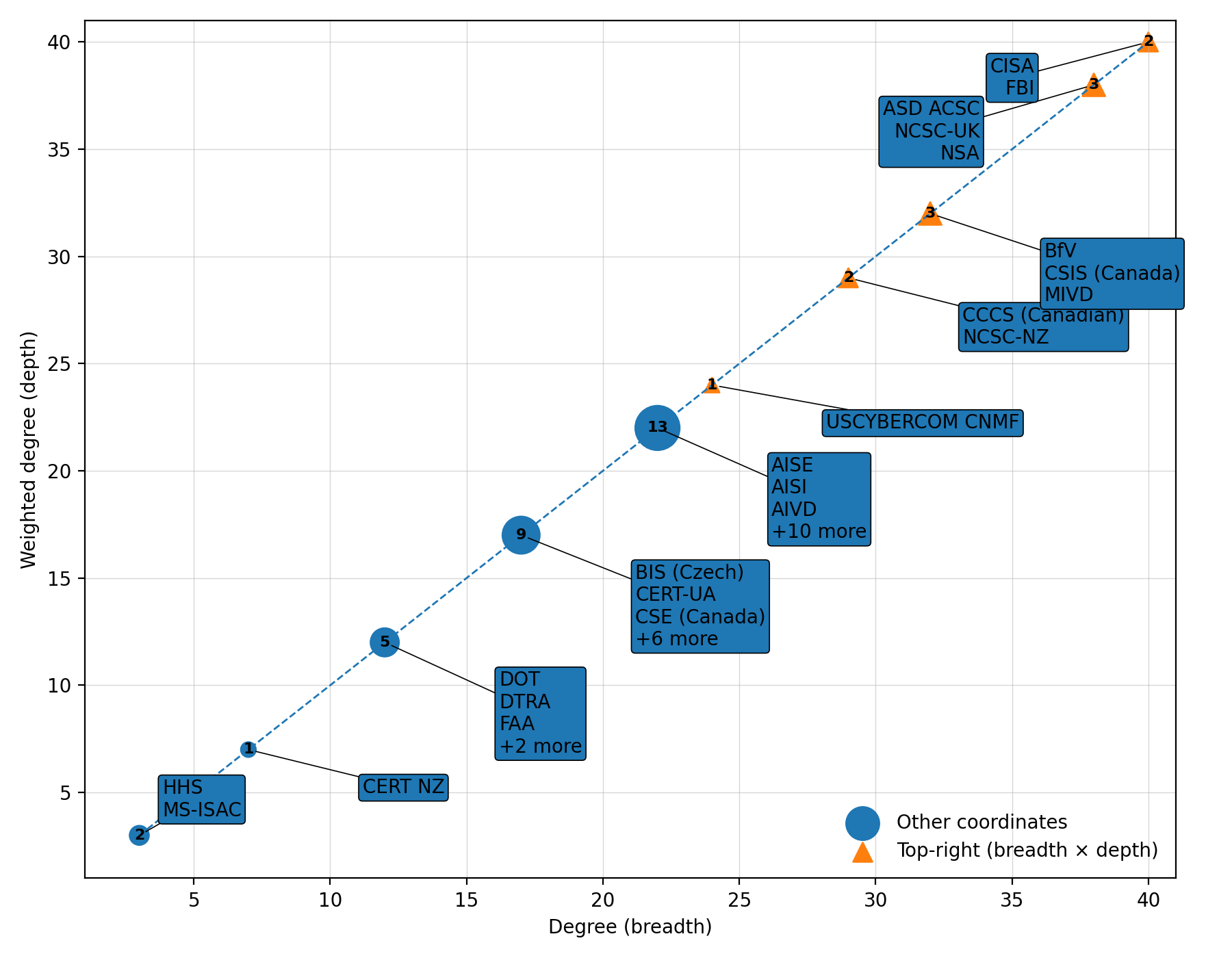}
\vspace{-6mm}
\caption{Agency collaboration landscape: Degree vs. Weighted Degree. Top-right agencies combine breadth and depth of collaboration.}
\vspace{-6mm}
\label{fig:scatter}
\end{figure}

\vspace{-2mm}
The scatterplot of degree versus weighted degree (Figure~\ref{fig:scatter}) illustrates the coalition’s core–periphery structure. We note that small visual compression effects cause agencies to appear closer to the diagonal in Fig.~\ref{fig:scatter} than their exact values in Table~\ref{tab:metrics} indicate; the underlying data correctly differentiate degree and weighted degree.
CISA and FBI sit at the extreme upper-right, combining both breadth (degree) and depth (weighted degree) of collaboration. NSA, ASD-ACSC, and NCSC-UK also cluster in this high–high region, underscoring their dual roles as hubs and repeat collaborators.  Agencies closer to the diagonal represent generalists, maintaining a balance between unique partners and repeated ties, while outliers above the diagonal would suggest strong-tie specialists.

It is important to note that the prominence of U.S. agencies is partly a function of data provenance, as the advisories originate from the CISA archive. While this reflects the vantage point of U.S.-released documents, the persistence of U.S. centrality across multi-nationally co-authored advisories suggests that the effect is not purely an artifact of collection but also indicative of their coordination role in global cyber defense. Future work should replicate this approach using advisories published by other national sources (e.g., NCSC-UK, CERT-IN) to test whether the network structure shifts under alternative points of origin.

What is surprising about the centrality measures of the co-authorship network is not the centrality of the US agencies (again, the documents were collected from CISA and as such are focused on the co-authorship of various other countries agencies with US security agencies); nor is it that partners like Australia and the United Kingdom have relatively high levels of centrality, as they are part of the Five Eyes intelligence cooperation community.  Rather, what stands out is that German domestic intelligence agency (BfV) and the Dutch military intelligence service (MIVD) actually have higher levels of centrality than some other Five Eyes intelligence partner agencies like Canada’s CSIS and CCCS, and New Zealands NCSC.  The closeness of Five Eyes cooperation on signals intelligence, of which cyber is an increasingly important component, has been well documented \cite{pfluke2019history}, but it may not be intuitive that the co-authorship network shows BfV and MIVD with such centrality.  Additional work will be needed with larger corpuses of documents to see whether such findings continue to appear.  This surprising centrality may also reflect the comparative advantage, specialization, or even quality of tradecraft of various countries intelligence agencies; for example the Dutch have invested heavily in cyber intelligence \cite{cerulus2021dutch}, and have garnered a reputation for being a serious player in intelligence despite being a relatively small country \cite{stein2024spy}.  
\vspace{-2mm}
\subsection*{Brokerage and Structural Vulnerabilities}
\begin{figure}[ht]
\centering
\includegraphics[width=0.75\textwidth]{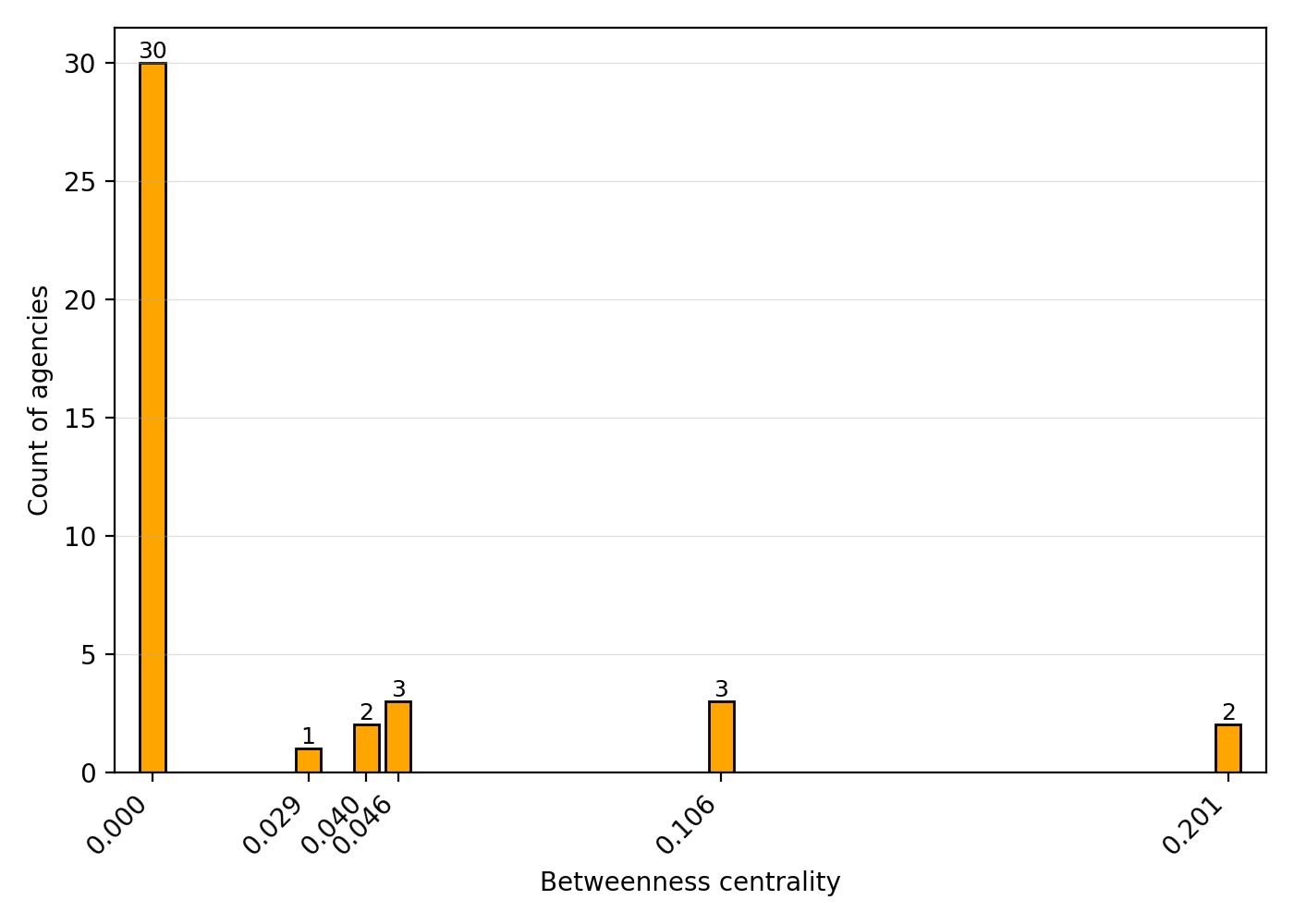}
\vspace{-6mm}
\caption{Distribution of betweenness centrality across agencies.}
\vspace{-6mm}
\label{fig:betweenness}
\end{figure}

\vspace{-2mm}
The distribution of betweenness centrality (Figure~\ref{fig:betweenness}) is highly skewed. 
Most agencies lie near zero, indicating they do not serve as unique bridges. 
A handful of hubs—CISA, FBI, NSA, ASD-ACSC, and NCSC-UK—stand out with markedly higher values. 
These agencies function as brokers connecting clusters, reducing the overall path length of the network. 
From a resilience perspective, this structure accelerates information flow but creates single points of failure: removing or disabling these brokers would disproportionately fragment the coalition.

\subsection*{Cohesion and Structural Nucleus}
\begin{figure}[ht]
\centering
\includegraphics[width=0.9\textwidth, height=.35\textheight]{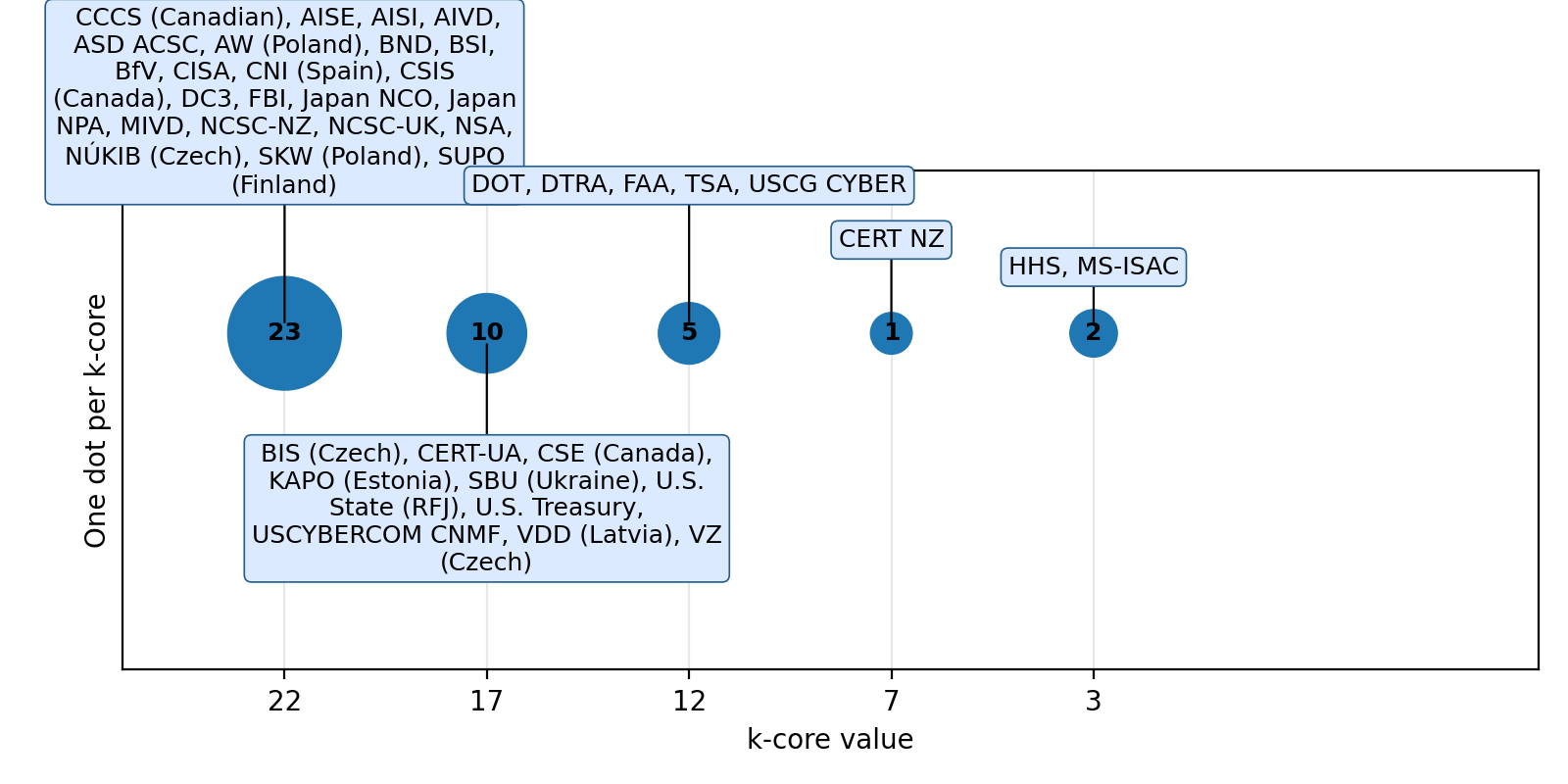}
\vspace{-6mm}
\caption{K-core decomposition of the CSA network. Agencies in the inner core are deeply embedded in the coalition backbone.}
\vspace{-6mm}
\label{fig:kcore}
\end{figure}

K-core analysis provides a complementary view of cohesion. 
Figure~\ref{fig:kcore} shows the distribution of agencies across nested shells. 
The innermost 22-core includes CISA, FBI, NSA, ASD-ACSC, NCSC-UK, and several European services, all tied together by dense and repeated co-authorship. Outside the 22-core, the next shells (e.g., 15-core, 10-core) largely consist of sectoral and regional agencies that participate more selectively; their presence illustrates a layered coalition, where outer shells can still reinforce resilience despite not being as densely interconnected as the innermost core. To be part of this core, an agency must be connected to at least 22 other members of that same core. Belonging to the 22-core indicates that the agency is embedded in a dense subnetwork where ties are both numerous and mutually reinforcing, signifying exceptionally high cohesion. Peripheral organizations such as HHS and CERT-NZ occupy outer shells, indicating more limited and selective participation. 
The presence of a large, dense core confirms the coalition’s robustness: even if one or two hubs are removed, the network retains cohesion through its embedded core.

\vspace{-2mm}
\subsection*{Synthesis}
\vspace{-2mm}
Taken together, these results portray a coalition that is both cohesive and asymmetric. 
The CSA network exhibits the hallmarks of a small-world system: high clustering, short paths, and a dense core. 
At the same time, the distribution of degree, betweenness, and k-core membership shows that resilience depends heavily on a small set of hubs, particularly CISA and FBI. 
The coalition is therefore robust against random loss of peripheral agencies but vulnerable to targeted disruption of its most central actors. 
This structural trade-off reflects a common tension in networked defense: efficiency and reach are achieved through concentration, but resilience requires deliberate reinforcement of secondary hubs.

\vspace{-2mm}
\section{Discussion}
\vspace{-2mm}

Our results reveal that the institutional network of Joint Cybersecurity Advisories (CSAs) is neither flat nor evenly distributed. 
Instead, it displays a hub-centric, small-world structure in which a small triad of U.S. agencies---CISA, FBI, and NSA---anchor the coalition, while key Five Eyes partners and select European services provide breadth and redundancy. The structure is efficient for information diffusion but vulnerable to over-reliance on a handful of brokers.

Adversaries targeting the credibility, operational capacity, or information systems of these hubs could disproportionately degrade coalition-wide coordination.  Also, this vulnerability need not be exploited by adversaries but could become salient because of the retreat of US agencies from pursuing certain cyber priorities; for example, recently the United States has defunded and dismantled much of the infrastructure and the programs it had built to support election cybersecurity \cite{cassidy2025cyber}.  Such retreat by central nodes in a co-authorship network could also render certain topics or types of coordination vulnerable.  Centrality is not just a measure that is useful for thinking about adversarial targeting, but also for thinking about organizational dependence and reliability.

The betweenness distribution highlights these vulnerabilities: CISA, FBI, and NSA serve as indispensable bridges whose elevated centrality indicates that many shortest paths in the coalition run through them. 
This accelerates coordination but also creates potential \emph{single points of failure}. 
Adversaries targeting the credibility, operational capacity, or information systems of these hubs could disproportionately degrade coalition-wide coordination. 
Redundancy at the k-core level partly mitigates this risk, but the skew in betweenness underscores the importance of developing \emph{surge capacity} across second-tier agencies such as ASD-ACSC and NCSC-UK.

The degree and weighted degree analysis further demonstrate that these hubs are not only widely connected but also repeatedly co-author advisories with the same partners, strengthening their brokerage role. 
High-degree agencies tend to connect to low-degree partners, forming star-like structures around CISA and FBI. 
This disassortative mixing mirrors the structure of cyber adversary networks themselves, such as botnets and ransomware affiliate programs, where central controllers coordinate loosely connected nodes. 
In this sense, the coalition’s structure is adversarially symmetric: distributed threats are met by a distributed defense anchored by powerful hubs. However, a key difference is network genesis. Adversary networks can form rapidly and organically around a specific criminal or state objective, while the defensive coalition is constrained by institutional mandates and diplomatic agreements. This may create an agility gap, a fruitful area for future research.
The asymmetry lies in governance: adversaries are not constrained by institutional boundaries, whereas defenders are. 
Future work should examine whether this structural tension limits the speed of response or the scope of information-sharing.

The community and modularity analysis add nuance to this picture. The CSA network shows modest modularity ($Q \approx 0.19$), suggesting that while national clusters exist, they are not deeply siloed. 
This indicates coalition members prioritize operational integration over national stovepipes in producing joint advisories. 
At the same time, modest modularity implies that meso-scale specialization---for example, health-sector agencies clustering together---is underdeveloped. This could imply that when a healthcare-specific ransomware campaign emerges (e.g., targeting hospitals), the response coalition is formed ad-hoc from the general pool of actors rather than from a pre-existing, tightly-knit health-cyber working group. Fostering such sector-specific communities could accelerate tailored responses.

\vspace{-2mm}
\section{Conclusion}
\vspace{-2mm}
This study provides the first network-science based mapping of U.S. and allied institutions co-authoring Joint Cybersecurity Advisories. Our findings highlight three core insights. 
First, the CSA network exhibits hallmark small-world properties: high clustering and short path lengths that enable rapid diffusion of alerts across a dense coalition. 
Second, it is hub-centric: CISA, FBI, and NSA dominate both degree and betweenness, acting as irreplaceable anchors but also presenting potential vulnerabilities. 
Third, the k-core analysis reveals a large structural nucleus including U.S. and Five Eyes partners, demonstrating redundancy and resilience, even if individual hubs are stressed or compromised.

For homeland security, these insights carry clear policy implications. 
Efforts to codify surge playbooks, expand international co-authoring frequency as a key performance indicator, and deliberately strengthen secondary bridges (e.g., ASD-ACSC, NCSC-UK, BfV) could mitigate reliance on the primary U.S. triad. 
Moreover, enhancing meso-scale community formation, particularly around critical infrastructure sectors, may increase adaptive capacity as threats evolve. 

From a methodological perspective, these analyses are not only retrospective but can also be extended predictively. 
Network embedding methods and anomaly detection algorithms could forecast shifts in coalition coordination patterns, anticipate vulnerabilities, and guide institutional interventions before adversaries exploit them. Dynamic graph neural networks could be trained on a time series of these CSA networks to predict which agencies are likely to form the core of a response to a novel, emerging threat vector, allowing for proactive resource allocation.

Ultimately, networks fight networks. 
Our results show that U.S. cyber defense collaboration is resilient yet hub-dependent, robust yet asymmetrically constrained. 
Recognizing and managing these trade-offs is essential for building coalitions capable of withstanding the distributed, adaptive, and adversarial networks they confront.

\vspace{-4mm}
\bibliographystyle{spmpsci}
\bibliography{refs}
\end{document}